\documentclass[9pt,preprint2,natbib209]{aastex}

\usepackage{dcolumn}
\usepackage{bm}
\shorttitle{ The diamagnetic phase transition in Magnetars}
\shortauthors{Wang et al.}

 \newcommand{\bq}{\begin{equation}}
 \newcommand{\eq}{\end{equation}}
 \newcommand{\bqn}{\begin{eqnarray}}
 \newcommand{\eqn}{\end{eqnarray}}
 \newcommand{\nb}{\nonumber}
 
\begin{document}

\title{ The diamagnetic phase transition of dense electron gas: astrophysical
applications }

\author{Zhaojun Wang\altaffilmark{1}, Guoliang L\"{u}\altaffilmark{1}, Chunhua Zhu\altaffilmark{1}, Baoshan Wu\altaffilmark{1}}
\email{xjdxwzj@sohu.com, guolianglv@gmail.com}
\altaffiltext{1}{School of Physical Science and Technology, Xinjiang
University, Urumqi, 830046, China}
\begin{abstract}
Neutron stars are ideal astrophysical laboratories for testing
theories of the de Haas-van Alphen (dHvA) effect and  diamagnetic
phase transition which is associated with magnetic domain formation.
The ``magnetic interaction'' between delocalized magnetic moments of
electrons (the Shoenberg effect), can result in an effect of the
diamagnetic phase transition into domains of alternating
magnetization (Condon's domains). Associated with the domain formation
are prominent magnetic field oscillation and anisotropic magnetic
stress which may be large enough to fracture the crust of magnetar
with a super-strong field. Even if the fracture is impossible as in
``low-field'' magnetar, the depinning phase transition of domain
wall motion driven by low field rate (mainly due to the Hall effect)
in the randomly perturbed crust can result in a catastrophically
variation of magnetic field. This intermittent motion, similar to
the avalanche process, makes the Hall effect be dissipative. These
qualitative consequences about magnetized electron gas are
consistent with observations of magnetar emission, and especially the
threshold critical dynamics of driven domain wall can partially
overcome the difficulties of ``low-field" magnetar bursts and the
heating mechanism of transient, or ``outbursting" magnetar.
\end{abstract}

\maketitle

\section{Introduction}
Degenerate Fermi electron gases occur in a huge variety of systems
for a wide range of density from non-relativistic metals to
relativistic neutron stars. Regardless of the particular substance,
there should appear to be quite similar over 
all magnetized systems in an applied or external magnetic field. During a
helical motion around the magnetic field, electron energy is quantized
into discrete Landau levels. If the successive Landau level interval
is higher than thermal energy of the system, $\hbar \omega_{\rm
c}\geq k_{\rm B}T$, electron gas can exhibit the nonlinear de Haas-van
Alphen (dHvA) effect. Here, $\hbar$ and $k_{\rm B}$ respectively are
the Planck and Boltzmann constants, $\omega_{\rm c}$ is the
cyclotron frequency. Whenever the Fermi energy approaches to the Landau
level, the magnetization oscillates rapidly with the field. The
oscillation is sinusoidal with a fundamental frequency determined by
an extremal area of the cross-section of the Fermi surface which is
normal to the applied magnetic field \citep{Lifshits1956}.

Using the impulsive field method, Shoenberg \citep{Shoenberg1962}
found an additional high amplitude in the second harmonic of the dHvA
oscillation, and proposed the concept of ``magnetic interaction''
among the conduction electrons. This interaction stems from the
overlap of helical orbits of electrons. For this reason, an electron
actually feels the magnetic induction (or magnetic flux density) $B$
instead of the magnetic field $H$. Usually, the difference between
$B$ and $H$ is $4\pi M$ and  small, where $M$ is the
oscillating part of magnetization (we use Gaussian units in this
paper). However, if the magnetic interaction is  large enough so that
the amplitude of magnetization oscillation can be comparable with
their period, the interaction may lead to a phase transition. This
transition belongs to a particular instability of electron gas,
the so-called diamagnetic phase transition \citep{Privorotskii1976}.
More accurately, when the differential magnetic susceptibility of
electron gas, $\chi_{\rm m}=\partial M/\partial B$, is greater than
$1/4\pi$, $\partial B/\partial H$ is negative and is not physical.
The stable state will have a spatial inhomogeneous   structure of
magnetic field with domains (i.e., Condon's domains
\citep{condon1966}) in which the magnetic induction and the
magnetization can take  one of two different values. There is a
transitional layer called domain wall (DW) between neighboring
domains with variation of magnetization. These magnetic domains have
been observed in some metals such as silver, beryllium, etc.
\citep{condon1966,Solt1996,Solt1999,Solt2000-1,Solt2002}. More
extensive review of theoretical and experimental progress about
the diamagnetic transition in metals can be found in
Gordon, Egorov, etc. \citep{Gordon2003,Egorov2010}. Magnetizing
process under a changing applied field would accompany DW
displacement. However, the motion is damped by the inhomogeneities
(e.g., crystalline imperfections) of the medium from which  typical
 dynamical critical phenomena (called depinning transitions) can
take place. All the essential elements of the depinning motion of
DW, as discovered in  experiments of the Barkhausen (BK) effect, are the
avalanche-like statistical properties of the magnetization noise:
distribution of size and duration of the avalanches, etc.
\citep{Alessandro1990(2),Narayan1996,Zapperi1998}.

Although all the theoretical and experimental conclusions as stated
above are limited to non-relativistic electron gases, we can extend
these results to neutron stars and their different subclasses, such as
magnetar especially. Magnetars, including Anomalous X-ray Pulsars
(AXPs) and Soft Gamma Repeaters (SGRs), are characterized by their
inferred dipolar magnetic field which is in the range of
$5.9\times10^{13}-1.8\times10^{15}$ G, and their long spin periods
 $5.2-11.8$ s. The most significant feature is the emission of
X-ray or $\gamma$-ray bursts and occasionally giant flares (with
duration of $0.1-10$ s, flux intensity of $10^{34}-10^{47}$ ergs
s$^{-1}$ at the peak) \citep{Woods2006,Mereghetti2008,Turolla2015}.
There is a consensus on the energy source of the persistent
and bursts emission powered by magnetic energy. Until now, however,
the emission mechanism remains an open question. In a framework of
starquake model, the evolution of the field imposes stress on solid
crust of neutron star, which is deformed elastically until
catastrophical rupture \citep{TD95}. However, whether the elastic
energy storing in crust could power the giant flares and the
``low-field" magnetars could been ruptured are still under debate
\citep{TD95,Rea10}. In the last few years, almost  all new
observed magnetars are transient magnetars marked by the
``outbursting" \citep{Ibrahim2004,Turolla2015}. This outbursting
mechanism has been explained as due to heat deposition into the star
surface, and then the limited region cools and shrinks \citep{Pons2012}.
Until now, the unambiguously heating mechanism has not been
identified.

The starquake model is strongly motivated by a similarity of
statistical distribution of SGR bursts and earthquakes
\citep{cheng1996,Gogus1999}. Such distribution belongs to
self-organized criticality \citep{Chen1991} in general. However, the
universal statistical aspects of SOC do not suggest that magnetar
bursts are necessarily from the crustquakes. Similar statistical
distributions have been found in solar flares \citep{Lu1993} and in
the BK effect of magnetic materials \citep{Durin2004}. Inside a
neutron star, the magnetic field configuration and matter
distribution under some suitable conditions can satisfy the
following three necessary conditions of the BK effect, i.e., the
existences of: (1) magnetic domains; (2) disordered defects; and (3)
a slowly changing magnetic field. There is no serious drawback about
the domain formation of a neutron star as having been proved in
\citep{Blandford82, Suh10,Wang2013} etc. {\bf \cite{Suh10}
considered the magnetization and  susceptibility of magnetar matter
for three different equations of state. They concluded that the
magnetic susceptibility can lead to the formation of magnetic
domains and a possible observable consequence of the SGRs bursts.
However, it has not yet been demonstrated how the magnetic domains
actually evolve and the disordered effect of the crust matter.}
Recent calculations about disordering defects (impurities,
dislocations, grain boundaries, etc.) in the crust of a neutron star
give a large impurity factor $Q\geq 10$
\citep{Jones2001,Horowitz2009} implying the presence of a very
strong magnetic hysteresis. At late times of the field evolution of
a neutron star (especially magnetar), the Hall effect of short
evolution timescale will dominate Ohmic decay \citep{Cumming2004}.
In spite of non-dissipation, the Hall effect can release magnetic
energy through a turbulent Hall cascade \citep{Goldreich92} which is
only or partially associated with the persistent emission of the
magnetar. However, after considering the damp and jerky motion of
DW, the Hall effect looked as if a continuous power provision source
can temporarily store the energy, which contributes a series of
sudden, incomplete releases of the accumulated energy. This behaves
like in a relaxation system of Palmer \citep{Palmer1999}. This
mechanism does not require absolutely crumbling of the crust and can
be regarded as a heating mechanism of the transient magnetar.

In this paper we consider particularly a degenerate
extreme-relativistic electron gas in the background of a crystal lattice
under a strong magnetic field. Except the magnetic interaction, we
neglect other complicated interactions of electrons with each other
or with the lattice viration. Under these conditions, the extension
of the dHvA effect and the diamagnetic phase transition to a relativistic
electron gas is simply the replacement of the cyclotron mass by the
relativistic mass. Using a very simple domain configuration, we
calculate the sizes of the domain thickness and  the surface tension
of the DW in its static state. Finally, we propose possible observational
effects of the diamagnetic transition, e.g., the magnetostriction
effect and the BK effect of dense matter.

The paper is organized as follows: In $\S$2, we review the general
theory of the dHvA effect and generalize the Lifshitz-Kosevich-Shoenberg
theory to relativistic electron gas of a neutron star. In $\S$3, we
give the conditions of the diamagnetic phase transition and the
corresponding phase diagram in a relativistic electron gas with
a spherical Fermi surface. In $\S$4, we examine the static structure
of domains and calculate the sizes of domain thickness and the
surface tension of DW. In $\S$5, we discuss the possible
observational applications to magnetars. Finally, Section $\S$6 is
devoted to our main conclusions and discussions.

\section{General theory of de Haas-van Alphen effect and application to dense matter}

For the completeness, we first briefly review the
Lifshitz-Kosevich-Shoenberg theory of magnetic oscillation in
metals. It is generalized to the case of a dense matter (such as
neutron star) with a similar Coulomb crystal lattice structure. A
detailed presentation about magnetic oscillation in metals is given
by Shoenberg in \citep{Shoenberg84}. For generalization
and application to a dense matter, we consider a typical neutron
star crust consisting of a neutron rich lattice in which the given
nucleus (with the nuclear charge $Z$) is present. Most of this crust
is in a radial shell of thickness about $10^{5}$ cm and total baryon
density, $\rho$, in the range $10^{6}-10^{14}$ g.$\rm cm^{-3}$.

\subsection{The theory of de Haas-van Alphen effect}

A realistic  system of conducting electrons in metal should be treated as
a Fermi liquid having complicated many-body interactions of
electrons (with e.g., each other, phonons or magnons). However, for
a magnetic oscillation theory about a system of approximately
independent quasi-particles, the behavior is determined by the
dispersion relation, $\varepsilon(\vec{k})$, an arbitrary dependence
of energy on wave vector $\vec{k}$, which is specified by the
periodic crystal potential. Acted on by a magnetic field, the
classical trajectory of a quasi-electron will be helical with a
cyclotron frequency, $\omega_{\rm c}$, given by
\begin{equation}
\omega_{\rm c}=\frac{2\pi eB}{c\hbar^{2}}/\left(\frac{\partial
s}{\partial \varepsilon}\right)_{\rm \kappa_{z}}=\frac{eB}{mc} \mbox{  .}
\end{equation}
where the cyclotron mass is defined as $m=(\hbar^{2}/2\pi)(\partial
s/\partial \varepsilon)_{\rm \kappa_{z}}$, $-e$ is the electronic
charge, $c$ is the velocity of light in vacuum, $\kappa_{z}$ is the
component of the wave vector along the magnetic field (in the z
direction), $s$ is the area of the cross section of the constant energy
surface at arbitrary $\kappa_{z}$ in the wave vector space.
Actually, the area $s$ is quantized and satisfies  the famous Onsager relation
\begin{equation}
s\left(\varepsilon,\kappa_{z}\right)=\left(r+\frac{1}{2}\right)\frac{2\pi eB}{c\hbar}\ \ \
\ \ \ \ \ \ \ \ (r=0,1,\cdots)\mbox{ .}
\end{equation}

At the ground state of an independent electron gas, all occupations
take place within the Fermi surface (FS). As the field increases,
the length of $\kappa_{z}$ with the largest area, which will
partially inside the FS, will shrink and vanish infinitely rapidly
when the area $s$ equals the extremal area $A$ of the cross-section of
the FS at $\kappa_{z}$. Such a special occupation happens
periodically with a fundamental frequency $F$
\begin{equation}
F=\frac{c\hbar}{2\pi e}A.
\end{equation}
For each of such occupations, the energy and magnetization experience
oscillations as the field varies. The oscillatory functions are
sinusoidal series with a fundamental frequency that can be
described by the extremal areas $A$ as Eq.(3). The absolute oscillation
amplitude is proportional to $|A^{\prime\prime}|^{-1/2}$ where
$A^{\prime\prime}=(\partial^{2}s/\partial \kappa^{2})_{\kappa=0}$.
Some corrections from finite temperature $T$, finite electron
relaxation time $\tau$ due to scattering, and electron spin can be
introduced independently as phase smearing that leads simply to a
multiplication factor for every harmonic term. If the separation of
successive Landau levels is larger than its thermal and
scatting energies, that is, $\hbar \omega_{\rm c}> k_{\rm B}T$ and
$\hbar \omega_{\rm c}>2\pi\hbar/\tau$, the oscillatory magnetization
per unit volume in the direction of the applying magnetic field is given by
\citep{Shoenberg84}
\begin{equation}
\begin{array}{ll}
M=&-(\frac{e}{c\hbar})^{3/2}\frac{2Fk_{\rm B}T}{(2\pi
B|A^{\prime\prime}|)^{1/2}}\\
&\sum_{p=1}^{\infty}\frac{\exp(-2\pi^{2}p\omega_{\rm e}/\omega_{\rm
c})\cos[p\pi(\Delta \varepsilon_{\rm s}/\hbar \omega_{\rm c})]}{p^{1/2}\sinh(2\pi^{2}pk_{\rm B}T/\hbar\omega_{\rm c})} \\
& \times \sin[2\pi p (\frac{F}{B}-\frac{1}{2})\pm \frac{\pi}{4}]
\mbox{     ,}
\end{array}
\end{equation}
where ``+"  or ``-" corresponds to the case of the minimum or
maximum extreme section respectively, $\omega_{\rm e}=2\pi/\tau$ is
the scatting frequency, $\triangle \varepsilon_{\rm s}/2$ is the
energy lifting due to the spin degeneracy. The
Lifshitz-Kosevich-Shoenberg formula of magnetization given above is
principally determined by the configuration of the constant energy
surface. It can be generalized to a relativistic
electron gas in a neutron star.

\subsection{Application of de Haas-van Alphen effect to dense matter}

Neutron star is one of the degenerate stars with electron and neutron
degenerations. Inside the deep crust of a neutron star, the electron gas
can be regarded as completely degenerate and extremely relativistic.
The Fermi energy $\psi_{0}$ in the absence of a magnetic field is
given by
\begin{equation}
\psi_{0}=(3\pi^{2})^{1/3}c\hbar n_{\rm
e}^{1/3}\approx51\rho_{12}^{1/3}Y_{\rm e}^{1/3}\mbox{ MeV,}
\end{equation}
where $n_{\rm e}$ is the number density of the electron gas and
$\rho_{12}$ (scaling as $10^{12}$ g $\rm cm^{-3}$) is the mass density
of the crust, $Y_{\rm e}$ is the number fraction of electrons.
Recent calculations of the Fermi energy can be found in 
\cite{Li2016}. Eq.(5) indicates that the Fermi energy is by far larger
than the rest-mass energy of the electrons, $\mu\equiv m_{\rm
e}c^2\approx 0.5 \rm MeV$ (where $m_{\rm e}$ is the rest mass of
electron), and much larger than the thermal kinetic energy $k_{\rm
B}T$ ($k_{\rm B}T<1$ KeV \citep{Pons07}). Considering the
lattice energy per electrons \citep{Coldwell1960}, $\varepsilon_{\rm
l}\approx 3.4\times10^{-3}Z^{2/3}\psi_{0}$, which is by far smaller
than the Fermi energy, we can treat the electron system more accurately as an ideal
gas.

A relativistic electron in an applied magnetic field is quantized with
the dispersion relation \citep{Jonhson49}
\begin{equation}
\varepsilon^2=c^{2}\hbar^{2}\kappa_{z}^{2}+\mu^{2}+\frac{c^{2}\hbar^{2}}{\pi}s=m^2c^4
\mbox{ ,}
\end{equation}
where $m$ is the relativistic mass of the electron, the quantized area
of cross section $s$ takes its values as those given in Eq.(2). From this dispersion
relation we find that the cyclotron frequency $\omega_{\rm c}$ is given
\begin{equation}
\omega_{\rm c}=\frac{2\pi eB}{c\hbar^{2}}/\left(\frac{\partial
s}{\partial \varepsilon}\right)_{\rm \kappa_{z}}=\frac{eB}{mc} \mbox{  .}
\end{equation}
This result indicates that the electron cyclotron mass is equal to it's
relativistic mass.

The maximum area of the cross section $A$ can be obtained from Eq.(6)
with $\kappa_{z}=0$ and $\varepsilon=\psi$ ($\psi$ is the chemical
potential of the electron gas in an applying field).
\begin{equation}
A=\frac{\pi}{c^{2}\hbar^{2}}(\psi^{2}-\mu^{2})\approx
\frac{\pi}{c^{2}\hbar^{2}}\psi^{2}\mbox{ ,}
\end{equation}
where we have supposed $\psi\gg \mu$ for the case of a neutron star.
The fundamental frequency $F$ of Eq.(3) relates to $B$ via
\begin{equation}
\frac{F}{B}\approx\frac{1}{2}\left(\frac{\psi}{\mu}\right)^{2}\frac{B_{\rm
Q}}{B}\mbox{ ,}
\end{equation}
where $B_{Q}=4.41 \times 10^{13}$ G is the quantum magnetic field of
the electron. The value of $|A^{\prime\prime}|=2\pi$ can be easily  obtained  from
Eq.(6).

The energy separation between successive $r$ at a fixed $\kappa_{z}$
is given by (when the quantum number $r$ is by far larger than 1)
\begin{equation}
(\triangle \varepsilon)_{\triangle r=1}=\left(\frac{\partial
\varepsilon}{\partial s}\right)_{\rm \kappa_{z}}(\triangle s)_{\triangle
r=1}=\hbar \omega_{\rm c}\mbox{  ,}
\end{equation}
which is formally similar  to the energy spacing in a no-relativistic
case except the rest mass now being replaced by the relativistic one at
the Fermi surface. Applying magnetic field
makes the spin degeneracy of the energy levels  lifted to,
\begin{equation}
\varepsilon=[c^{2}\hbar^{2}\kappa_{z}^{2}+\mu^{2}+(2r+1+s')\mu
\epsilon _{\rm c0}]^{1/2}\mbox{ ,}
\end{equation}
where $s'=\pm 1$ and $r=0,1,\cdots$, are the  spin and Landau
quantum numbers of electron respectively. $\epsilon _{\rm
c0}=(eB/m_{\rm e}c)$ is the cyclotron energy of an electron in
the non-relativistic. Near the FS, the energy splitting due to electron
spin is small and can be expanded in Taylor series. To the first order,
the energy spilling is
\begin{equation}
\varepsilon\approx \psi\pm
\frac{1}{2}\triangle\varepsilon\approx\psi\pm\frac{1}{2}\hbar
\omega_{\rm c}\mbox{ ,}
\end{equation}
where ``-" or ``+" corresponds to the spin-up and spin-down
respectively. The reduction factor of
$\cos(p\pi\triangle\varepsilon/\hbar\omega_{\rm c})$ in Eq.(4)   due
to the spin is equal to $(-1)^{\rm p}$.

Keeping only the lowest-order term, the final form of Eq.(4) is
given by
\begin{equation}
4\pi b M=a(T,T_{\rm D},H_{\rm 0})\sin[b(h+4\pi M)]\mbox{ ,}
\end{equation}
where $H_{\rm 0}=B_{\rm 0}$ is the magnetic field of the center of
an oscillation cycle, $h=H-H_{\rm 0}$ denotes the deviation of the
magnetic field $H$ from $H_{\rm 0}$ (supposing $h\ll H_{\rm 0}$),
$b=2\pi F/B_{\rm 0}^{2}$. In Eq.(13), we have neglected the constant
phase. The reduced amplitude of magnetization due to thermal
temperature, Dingle temperature $T_{\rm D}$ (or the relaxation time
of the electron scatting, $\tau=2\pi \hbar/kT_{\rm D}$) is given by
\begin{equation}
a(T,T_{\rm D},B_{\rm 0})=a_{\rm 0}(B_{\rm 0})\frac{\lambda T}{\sinh
(\lambda T)}\exp(-\lambda T_{\rm D})\mbox{ ,}
\end{equation}
\begin{equation}
a_{\rm 0}(B_{\rm
0})=\frac{\alpha}{\pi}\left(\frac{\psi}{\mu}\right)^{3}\left(\frac{B_{\rm Q}}{B_{\rm
0}}\right)^{3/2}\mbox{ ,}
\end{equation}
where $\alpha$ is the fine structure constant,
$\lambda=2\pi^{2}k_{\rm B}/\hbar \omega _{\rm c0}$ and $\omega _{\rm
c0}$ takes the value of Eq.(7) with $B=B_0$. This result is formally
similar to the one given in metals \citep{Gordon1999}, and also similar to
the result presented in \citep{Wang2013}, in which the temperature
and scatting effects were neglected.

Under a strongly quantizing magnetic field which can be prevailed in
the magnetar interior or the most out-crust, only a few Landau levels
are occupied. Then, the above theory and calculations about the electron gas
magnetization cannot be applied.
In this case, 
matter behavior and the equation of state at the thermodynamical
equilibrium or the nonequilibrium $\beta$-process have been studied
in \cite{Lai1991,Chamel2012,Basilico2015,Wang14}, etc.
In  \cite{Wang14}, we have analytically calculated the
magnetization of the electron gas acted with strongly quantizing
magnetic field. The magnetization exhibits also the similar
oscillatory behavior as that described by Eq.(4).

It should be noted that  the Lifshitz-Kosevich-Shoenberg theory to a neutron star is applicable
only in  simple models of independent electrons. The finite
temperature effect is treated as only blurring out slightly the
boundary of the occupation and the un-occupation, and within a finite
relaxation time only blurring out the sharpness of quantized states.
The general magnetization theory of an electron system should consider
the many-body interactions (with each other or with phonons or
magnons). A quite general result of the interactions is the
modification of the dispersion relation $\varepsilon(\vec{k})$ of the
independent particle model. Firstly, the independent particle energy
is shifted to a dynamical quasi-particle energy with an additional
term , $\varepsilon=\varepsilon(\vec{k})+
\triangle(\varepsilon-\psi)$. Secondly, the energy is broadened
according to the Lorentzian fashion characterized by a parameter
$\Gamma(\varepsilon-\psi)$, where $\psi$ is chemical potential of the
electron system. These modifications can be obtained by quoting a basic
concept of the self energy which is a complex quantity denoted by
$\sum(\varepsilon-\psi)=\triangle(\varepsilon-\psi)-i\Gamma(\varepsilon-\psi)$.
As Shoenberg had showed \citep{Shoenberg84}, the
Lifshitz-Kosevich-Shoenberg formula of an independent electron system
remains valid, but the parameters have to be modified
appropriately. The basic theory about the effect of many-body
interactions is beyond the scope of this paper (for detailed
discussions of electron-electron and electron-phonon interactions
readers are referred to \cite{Luttinger1960} and \cite{Engelsberg1970}).

\section{The diamagnetic phase transition of relativistic
electrons inside a neutron star}

Eq.(13) is the  correct form of the Lifshitz-Kosevich formula after
Shoenberg's magnetic interaction is taken into account. In its
derivation, we have implicitly assumed that the magnetization is a
function of the magnetic induction, i.e., $M=M(B)$. This suggests
that the magnetization depends also on the  ordering of the orbital
magnetic moments of electrons. Usually, the difference between $B$
and $H$ is very small. If the amplitude of the magnetization
oscillation is comparable with their period,  however, the
interaction can lead to the diamagnetic phase transition. In this
case, the self-consistent solution of Eq.(13) about the
magnetization is multiple-valued. In fact, as discussed in
\cite{condon1966}, the material will separate into two phases with
parallel and anti-parallel magnetizations. This transition is
different from the ferromagnetic transition. In the former  the
magnetic ordering comes from the classical magnetic interaction,
while in the latter it is from the quantum exchange interaction. The
diamagnetic phase occurs in each cycle of the dHvA oscillations
whenever the reduced amplitude of the oscillations is equal to 1,
$a(T_{\rm d},T_{\rm D},B_{\rm 0})=1$, where $T=T_{\rm d}$ is the
critical temperature.

One of the necessary conditions of the diamagnetic phase transition is
$a_{0}(B_{\rm 0})>1$ given in Eq.(15). According to $\psi_{0}$ of Eq.(1),
we can approximately take $\psi\approx\psi_{0}$ in spite of that the
chemical potential of an ideal electron gas slowly declines with the
increase of the magnetic field \citep{Ingraham1987}. From Eq.(15), the
condition is
\begin{equation}
B_{\rm 0}<\left(\frac{\alpha}{\pi}\right)^{2/3}\left(\frac{\psi_{0}}{\mu}\right)^{2}B_{\rm
Q}\mbox{ .} \label{eq:bbq}
\end{equation}
{\bf If we take a value of $\psi_{0}\sim25$ MeV inside the deep
crust,
 Eq.(\ref{eq:bbq}) becomes $B_{\rm 0}<44B_{\rm Q}$ (the rest
energy of an electron is about $\mu\sim0.5$ MeV). Almost all of the normal
neutron stars including magnetars satisfy this condition. While some
SGRs with a magnetic field near $100B_{\rm Q}$ (e.g. SGR 0526-66, SGR
1806-20, SGR 1900+14), only the diamagnetic phase transitions occur
in the liquid core or/and inner crust as sketched in Fig.1.}

The condition of a large Landau quantum number requires $\psi_{\rm
0}\gg \hbar \omega_{\rm c}$. From the definition of $\omega_{\rm c}$
in Eq.(7), this condition is satisfied  as long as
\begin{equation}
\psi_{\rm 0}\gg \mu\left(\frac{B_{\rm 0}}{B_{\rm Q}}\right)^{1/2}\mbox{  .}
\end{equation}
{\bf Except for some very particular cases of low Landau quantum
number occupations, the Lifshitz-Kosevich-Shoenberg theory is
adequate. Actually, \cite{Shoenberg84} suggested the largest quantum
number $r_{\rm max}> 2$. If $r_{\rm max}\leq 2$, the diamagnetic
phase does not really occur in spite of the similar oscillation
property of Eq.(4) \citep{Wang14}.}

\cite{Chamel2012} has given the out crust structure and composition
of a neutron star based on experimental data of atomic mass and
complemented with theoretical prediction of the
Hartree-Fock-Bogoliubov method \citep{Goriely2010}. Their results
only included the effect of the Landau quantum of the electron gas in a strong
magnetic field. However, the analysis of \cite{Basilico2015}
accounted for the influence of the field on the nuclear binding
energy and showed that the binding energy does not increase by more than
10\%. The predicted composition in the two cases is similar, and the main
difference is in the most outer crust \citep{Basilico2015}. {
Using the  results of the outer crust structure and combining those
of \cite{Chamel2012} at zero-temperature, the possible
phase transition regions ($a_{\rm 0}>1$) are indicated in Fig.1.}

\begin{figure}
\centering
  \includegraphics[scale=0.3, angle=-90]{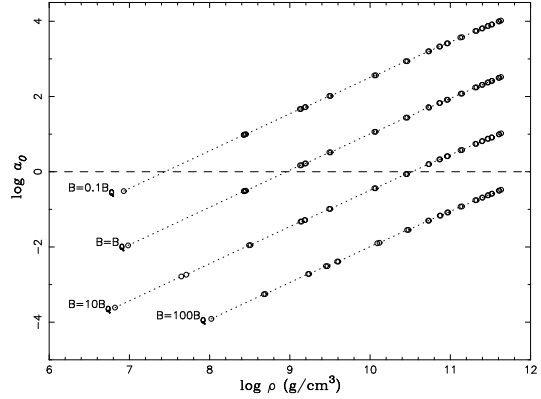}
  \caption{The largest amplitude of the dHvA oscillation $a_{0}$ in Eq.(15)
  versus neutron star densities inside the outer crust below neutron drip.
  The circles refer to numerical calculations of \cite{Chamel2012}
  for various compositions with different magnetic field
  ($B=0.1B_{\rm Q},B_{\rm Q},10B_{\rm Q},100B_{\rm
  Q}$). Long-dashed line ($a_{0}=1$) divide no (below line) and possible
  (above line) phase transition regions.}

\end{figure}

The effects of finite temperature $T$, finite electron relaxation
time $\tau$ due to the impurity scattering, had been introduced
independently as phase smearing that lead to a multiplication factor
$\lambda T/\sinh(\lambda T)$ and $\exp(-\lambda T_{\rm D})$ in
Eq.(14) respectively. However, as shown in standard textbooks \cite
{Landau80}, there are other necessary conditions of the magnetic
oscillation restraining the temperature and electron relaxation. If
the separation of successive Landau levels near the Fermi surface is
not larger than the thermal kinetic and scatting energies, $\hbar
\omega_{\rm c}< kT$ and $\hbar \omega_{\rm c}<kT_{\rm D}$, the
diamagnetic phase does not occur. The Dingle temperature from
impurity \citep{Itoh1993} is
\begin{equation}
T_{\rm D}\approx 3.3\times 10^{6}\Lambda_{\rm
eQ}(\frac{\epsilon}{100})(\frac{Z}{40})^{-1}(\frac{Q}{10})\\\ {\rm
K},
\end{equation}
where $\Lambda_{\rm eQ}$ is the Coulomb logarithm of order
unity, $\epsilon=\psi/\mu$, and $Q$ is the impurity factor which is
defined as the mean square charge deviation $\langle (\triangle
Z)^{2}\rangle$. Here we have scaled the impurity $Q$ by a large
value $Q\sim10$ \citep{Jones2001}.

Combining all these necessary conditions given above with the critical
condition of $a(T_{\rm d},T_{\rm D},B_{\rm 0})=1$, the phase
transition curve can be determined from the following equations
\begin{equation}
T_{\rm d}=0\ \ \ (\hbar \omega_{\rm c}\leq k_{\rm B}T_{\rm D})\mbox{
,}
\end{equation}
\begin{equation}
T_{\rm d}=\hbar \omega_{\rm c}/k_{\rm B}\ \ \ (k_{\rm B}T_{\rm
D}<\hbar \omega_{\rm c}\leq k_{\rm B}T_{\rm d})\mbox{  ,}
\end{equation}
\begin{equation}
a_{\rm 0}\frac{\lambda T_{\rm d}}{\sinh[\lambda T_{\rm
d}]}\exp[-\lambda T_{\rm D}]=1 \ \ \ (\hbar \omega_{\rm c}>k_{\rm
B}T_{\rm d})\mbox{.}
\end{equation}
{\bf The  range of the temperature of the phase transition in the
outer crust of a neutron star with different magnetic fields are
showed in Fig.2. The regions between the phase transition curves and
the abscissa axis indicate the possibility of the diamagnetic phase
transition. Figs.3 and 4 are the phase diagrams which show the
phase-transition temperature as a function of the magnetic field for
different matter densities. The value of the Fermi energy in Fig.3
is taken to be 25 MeV for strontium inside the deep outer crust,
while in Fig.4 it is 2.5 MeV for iron envelope. The Dingle
temperature, as a parameter in Figs.3 and 4, is taken as $Q=10,100$
in Eq.(18). The results indicate that the effect of the Dingle
temperature is not obvious in the  circumstance considered. }
\begin{figure}
\centering
  \includegraphics[scale=0.3, angle=-90]{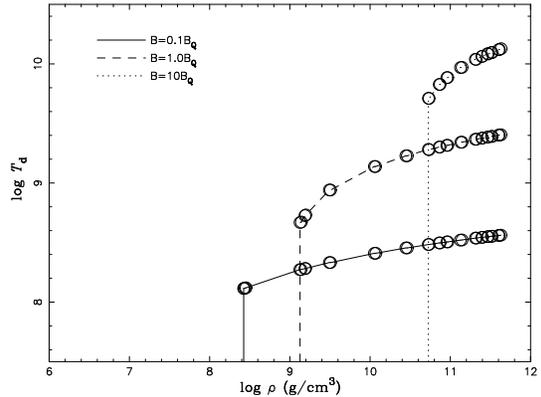}
  \caption{Phase diagrams for the diamagnetic phase transition in
the outer crust. The phase-transition temperature is  a function
of matter densities, and the magnetic field and  Dingle temperature
($Q=10$) are choosen according to Eqs.19-21. The circles refer to
the numerical calculations of \cite{Chamel2012} for various compositions
with different magnetic fields.}

\end{figure}
\begin{figure}
\centering
  \includegraphics[scale=0.3, angle=-90]{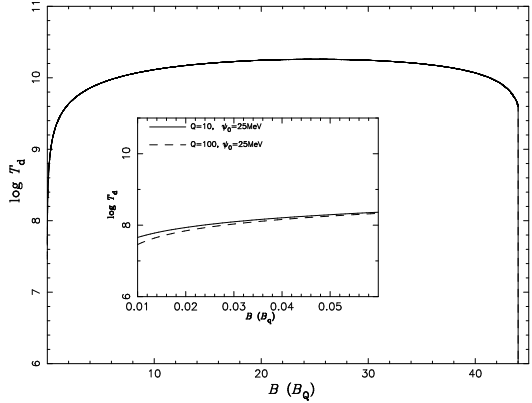}
  \caption{Phase diagrams for the diamagnetic phase transition versus
  the magnetic field in the deep outer crust ($\psi_{0}=25$ MeV). The Dingle temperature
  $Q=10$ (solid line) or $Q=100$ (dashed line) is taken as a parameter.
  The insert shows a distinguishable region for different Dingle temperatures.}

\end{figure}
\begin{figure}
\centering
  \includegraphics[scale=0.3, angle=-90]{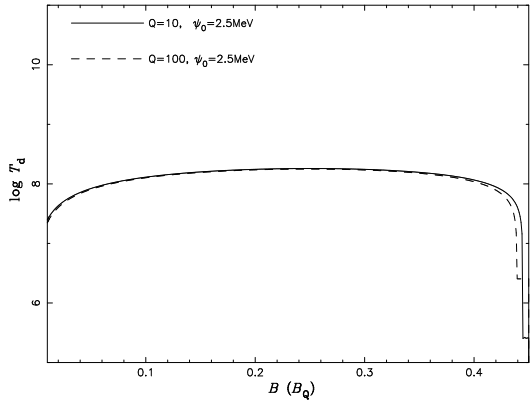}
  \caption{Phase diagrams for the diamagnetic phase transition versus
  the magnetic field in the deep outer crust ($\psi_{0}=2.5$ MeV). The Dingle temperature
  $Q=10$ (solid line) or $Q=100$ (dashed line) is taken as a parameter.}

\end{figure}

Taking Eq.(13) at $h=0$ and expanding the right-hand side to the
third order on $M$, we obtain an equation of order parameter, i.e.,
the magnetization in the Condon domains \citep{Gordon1999}.
\begin{equation}
\frac{M_{\rm s}}{B_{0}}=\pm 5\times
10^{-5}\sqrt{\frac{3(a-1)}{2a}}\left(\frac{\epsilon}{100}\right)^{-2}\frac{B_{0}}{B_{\rm
Q}}.
\end{equation}
Eq.(22) indicates that the order parameter approaches to zero when the
reduced amplitude $a(T,B_{\rm 0},T_{\rm D})$ tends to 1. Near the
critical point, the order parameter is proportional to $(T_{\rm
d}-T)^{1/2}$, showing the general characteristics of
a continuous phase transition. ``This fact is in agreement with the
long-range character of magnetic interactions in the cooperative
system of orbital magnetic moments of the electron gas"
\citep{Gordon1999}. Far from the critical point, the fraction of the
saturation magnetization, $M_{\rm s}/B_{0}$, is proportional to the
central magnetic field at a fixed Fermi energy. Near the deepest outer
crust, for example, $\epsilon \approx50$ for $B_{*}=0.1$ and
$B_{*}=100$ in table-4 and table-3. The fractions are approximately
of $2.4\times10^{-5}$ and $2.4\times10^{-2}$, respectively,
indicating the potential importance of magnetizing in magnetars.

\section{The Condon domain structure inside neutron star}

The diamagnetic phase occurs in each cycle of the dHvA oscillation with
a period determined by the magnetic field and Fermi energy of the electron
gas. In the  interior of the solid crust of a neutron star, both magnetic field
and Fermi energy change on the depth from the crust surface. Thus,
the possible phase transition regions are arranged in layers. In the
layer structure, domain arrangement with alternate magnetization is
energetically favorable by reducing the demagnetization energy as
discussed by Condon \citep{condon1966}. Moreover, decreasing the
domain width (meaning,  increasing the number of domains) can reduce
further the magnetostatic energy because of the  rapidly falling off of
the free magnetic poles. However,  this is not favorable due to
the increase of  the magnetostatic energy of DWs. Minimizing the sum of
these two contributions leads to an optimum domain size. Approximate
calculations of the size of the simplest domain structure will be
given below. For every domain layer of a neutron star, we
first calculate the thickness and surface tension of DW.

\subsection {The thickness and surface tension of DW}

We start with the assumption that the wall thickness is of order of
$d_{\rm c}$, the cyclotron diameter. Because of the inhomogeneous
magnetization in this thickness, there is an additional free energy
term in the thermodynamic potential density. The additional term is
proportional to $(d\hat{m}/dx)^{2}$ where $\hat{m}=4\pi b M$ is the
dimensionless magnetization. Correspondingly, the thermodynamic
potential density takes the form \citep{Privorotskii1976}
\begin{equation}
\Omega=\frac{1}{4\pi
b^{2}}\left[a\cos(bh+\hat{m})+\frac{1}{2}\hat{m}^{2}+\frac{1}{2}K\left(\frac{d\hat{m}}{dx}\right)^{2}\right]
\mbox{ ,}
\end{equation}
where $K$ is the positive coefficient of the inhomogeneous term
which is proportional to $d_{\rm c}^{2}$. The dynamics of
magnetizing in a non-zero external magnetic field is described by
the Landau-Khalatnikov equation, $\partial M/\partial t=-\Gamma \delta
\Omega/\delta M$, where $\Gamma$ is the dynamic coefficient. For the
potential density given by Eq.(23), we find, 
\begin{equation}
\frac{\partial \hat{m}}{\partial t}=-4\pi
a\Gamma\left[-\sin(bh+\hat{m})+\frac{\hat{m}}{a}-\frac{K}{a}\frac{\partial^{2}\hat{m}}{\partial
x^{2}}\right] \mbox{ .}
\end{equation}
This equation is ascribed to the nonlinear reaction diffusion
equations. The static form of Eq.(24) has a kink solution. The
thickness of a static DW is defined as
$D=2M_{s}/(\frac{dM}{dx}|_{x=0})$ which can be estimated as that given in
 \cite{Bakaleinikov2012}. There are two limiting cases
of $a$, one is that  $a$ is greater than and close to 1, and the other is that
$a\gg 1$. For the first case, the thickness is given by
\begin{equation}
D=6.7\times
10^{-8}\frac{1}{\sqrt{a-1}}\left(\frac{\epsilon}{100}\right)\left(\frac{B_{0}}{B_{\rm
Q}}\right)^{-1} \mbox{cm,}
\end{equation}
and for the second case it is
\bqn
D&=&7.4\times
10^{-8}\left[1+\left(\frac{\pi^{2}}{8}-1\right)/a\right]\nb\\
&&\times
\ \left(\frac{\epsilon}{100}\right)\left(\frac{B_{0}}{B_{\rm
Q}}\right)^{-1} \mbox{cm,}
\eqn
where we have taken $K=d_{\rm c}^{2}/4$. In the deep crust of
a neutron star, the reduced effect of the oscillation magnitude due to
temperature and scatting is neglected and $a\gg 1$. From Eq.(26),
the thickness is the order of $D_{0}\approx 0.5\pi d_{\rm c},$. This
is indeed true, as  the DW is order of the electron cyclotron diameter.

The surface tension $\sigma$ of the DW, defined as
$\sigma=\frac{1}{2}K\int_{-M_{\rm S}}^{M_{\rm S}}\frac{dM}{dx}dx$,
can be calculated from the kink solution of the static equation of
Eq.(24). As a function of deviation from the transition critical
point $a-1$, the tension is
\begin{equation}
\sigma=4.8\times
10^{22}\frac{(a-1)^{3/2}}{a}\left(\frac{\epsilon}{100}\right)\left(\frac{B_{0}}{B_{\rm
Q}}\right)^{-1}\ \rm{erg}\ \rm{cm}^{-2}.
\end{equation}
Thus, the thickness of Eq.(25) and the surface tension of Eq.(27) for
the DW indicate the dependence of $(T_{\rm d}-T)$, which is typical
to the mean-field theory of  continuous phase transitions. This is
the reflection of the magnetic interactions leading the transition.

Furthermore, from Eqs.(25-27) we can quantify the thickness and
surface tension with the outer crust structure and compositions for
different magnetic fields in Tables 1-4. But we can get the
qualitative properties of the thickness and tension by combining
Eq.(14). In the deep outer crust with a fixed Fermi energy and $a\gg1$,
the thickness and tension are inversely proportional to $B_{0}$ and
$B_{0}^{7/4}$, respectively. The result implies that a magnetar is
relatively easy to experience depinning transitions than a normal
neutron star because a magnetar is thinner and has a low tensional DW
(which will be discussed in section 5.2).

The above calculations of the thickness and surface tension of DW do not
involve the elastic energy of the crystal background given  in Eqs.(23) and
(24). These results underestimate the actual thickness and
overestimate the surface tension. Probably, the cyclotron diameter
plays the role of the lower limit of the DW thickness.

\subsection{The Condon domain size}

For the simplest domain structure with parallel and anti-parallel
(along z direction) magnetizations, the sizes of the domain thickness
(along x direction) $\delta X$ and the domain (or DW) width $\delta
Z$ can be calculated approximately. The size of DW has been given in
Eq.(25) or Eq.(26). In each layer of the transition where the
magnetic field is assumed to be a constant, the Fermi energy
increases with the depth below the surface of the neutron star. The
variation of the Fermi energy corresponding to the phase difference
$2\pi$ of the sinusoidal argument in Eq.(13) determines the
thickness of domains and DWs. Provided that the surface gravity is
supported by the degenerate electron pressure, the thickness is
similar to the result of \cite{Blandford82}
\begin{equation}
\delta Z\approx 50g_{\rm 14}^{-1}Y_{\rm
e}\left(\frac{\epsilon}{100}\right)^{-1}\frac{B_{\rm 0}}{B_{\rm Q}} \mbox{ cm,}
\end{equation}
where the surface gravity is $g=10^{14}g_{\rm 14}$ cm $\rm s ^{-2}$.
This result indicates the thickness of the domain and DW is
proportional to the magnetic field at a fixed region. For a fixed
magnetic field configuration the thickness is reverse proportional to
the electron Fermi energy. At the deep outer crust of a magnetar with
the Fermi energy 25 MeV and the magnetic field 10$B_{\rm Q}$, the
computed thickness is about 306 cm.

The reason of the existence of the Condon domain is that the magnetostatic
energy is greatly reduced. This magnetostatic energy comes
dominantly from the domains and the DWs. First we consider
magnetostatic energy of the simplest domain structure in which there
are many plate-like domains with a magnetization $M_{\rm s}$ along the
$+z$ or $-z$. Bearing in mind that the domain thickness in the
$z$-direction is larger than that in the $x$-direction, we can
simplify the domains by extending indefinitely in the $-z$ direction.
Using the method of magnetic potential which is induced by the
magnetic free poles at the surface of the transition layer, the
magnetostatic energy can be calculated. This calculation was carried
out by \cite{Kittle1949} and the magnetostatic energy per unit area
of the surface $\varepsilon_{\rm m}$ is $\varepsilon_{\rm
m}=0.85M_{\rm S}^{2}\delta X$. In fact, on the top and  bottom of
the surface there exits same free poles with opposite signs, so that
the magnetostatic energy is given by two times of the result. On
the other hand, the total area of DWs per unit area of the surface
is $\delta Z/\delta X$. The equilibrium domain thickness is given by
minimizing the total energy, $\varepsilon_{\rm m}=1.7M_{\rm
S}^{2}\delta X+\sigma \delta Z/\delta X$. Solving
$\frac{d\varepsilon_{\rm m}}{d\delta X}=0$, we have the size of the
domain thickness, $\delta X= 0.77\sqrt{\sigma \delta Z}/M_{\rm S}$,
which is in the same way to be proportional to the geometric average of the
domain width and the DW thickness as in ferromagnetic materials in
spite of different phase transition mechanisms. Combining
Eqs.(22),(27) and (28), the size is
\bqn
\delta X&=&4.2\times10^{2}Y_{\rm
e}^{1/2}g_{14}^{1/2}(a-1)^{1/4}\left(\frac{\epsilon}{100}\right)^{3/2}\nb\\
&& \times \left(\frac{B_{0}}{B_{\rm
Q}}\right)^{-3/2}\mbox{cm.}
\eqn 
At the deep outer crust of the magnetar with the Fermi energy 25 MeV and
the magnetic field 10$B_{\rm Q}$, the computed domain size along the x
direction is about 4.4 cm,  which is actually less than the domain
thickness of 306 cm.

All of these results indicate that the outer crust of a neutron star
consists of  discrete Condon domain laminas whenever  the diamagnetic phase
of the electron gas occurs. The size of each domain lamina is
given roughly by Eqs.(28) and (29).

\section{Possible observational effects in magnetars}

With the natural presences of a very large Fermi energy, strong or
super-strong magnetic field, relatively low temperature and Dingle
temperature, the diamagnetic phase can occur unprecedentedly inside
the outer crust of a neutron star. Available magnetic free energy
increases rapidly as $B_{0}^{4}$ because of the nonlinear
magnetization. Except the discrepancy of the interaction mechanism,
the diamagnetic phase is qualitatively the same as  the
ferromagnetic phase. Many observational effects in ferromagnetic
materials such as magnetostriction, magnetocaloric
\citep{Gschneidner2005} and Barkhausen effects etc., should have
their counterparts in a neutron star. Especially, the crust fracture
due to the sharp magnetostriction or the intermittent motion of DWs
can release elastic energy and the magnetic free energy deposited in
the outer crust. This released energy can provide (or partially) the
observable emissions from magnetar.

\subsection{Magnetostriction effects}

Accompanying the Landau quantization, all kinds of thermodynamic
quantities such as the thermodynamic potential density of Eq.(23)
and its derivative, the temperature and specific heat, the electron
Fermi energy etc., oscillate  depending on the field variation.
This variation of the Fermi energy results from the oscillatory
field dependence of the volume of the electron gas under a fixed
pressure. In general the formation or adjustment of domain structure
involves anisotropic magnetostrictive stress. The magnetostriction
will be balanced by elastic stress in the lattice of the solid crust,
provided that it was beyond the yield stress. Having considered the
additional contribution of lattice stress energy to the
thermodynamic potential density of the outer crust system, the
lattice strain can be determined by its derivative with respect to
the stress. If explicitly taking  the elastic energy into account, the
generalized thermodynamic potential density $\Omega^{\prime}$ of
Eq.(23) can be expressed  in terms of the stress tensor $\sigma$ in the form,
\begin{equation}
\Omega^{\prime}=\Omega-\frac{1}{2}(S_{ikpq})_{0}\sigma_{
ik}\sigma_{pq}.
\end{equation}
In the absence of the magnetic field, here the elastic compliance tensor is
denoted by $(S_{ikpq})_{0}$. Therefore, the second term on the right-hand
side of Eq.(30) is only the elastic energy of the lattice
stressed by the external stress. Due to electrical neutrality, the
deformation of electron gas is identical with the lattice strain.
The first term on the right hand of Eq.(30) includes the
thermodynamic potential density of electron gas, in which the Fermi
energy (or the oscillatory frequency) can change as changing the
field. Corresponding to  a given stress $\sigma_{ik}$, the strain is,
$\varepsilon_{ik}=-\partial \Omega^{\prime}/\partial
\sigma_{ik}=\widetilde{\varepsilon}_{ik}+(S_{ikpq})_{0}\sigma_{pq}$.
Here, we are only interested in the term of oscillatory
magnetostriction, $\widetilde{\varepsilon}_{ik}=-\partial
\Omega/\partial \sigma_{ik}$, which is given approximately by
$\widetilde{\varepsilon}_{ik}=-(\partial\ln F/\partial
\sigma_{ik})MB_{0}$. A more detailed discussion about this result
can be found in \cite{Shoenberg84}. For a spherical
Fermi surface of free electrons, its volume is inversely
proportional to the gas volume $V$ as varying the stress. For this
simplification and the linear relation of $F$ to the area $A$ (see
Eq.(3)), we have $\partial\ln F/\partial
\sigma_{ik}=-\frac{2}{3}(\partial\ln V/\partial \sigma_{ik})$. If
neglecting the anisotropy of the compressibility coefficient and the
Poisson ratio, we obtain $\partial\ln F/\partial \sigma_{ik}=-G/3$, where $G$
is the shear modulus.

In the deep crust of a neutron star, the lattice rigid with respect to
the terrestrial standard is enormous with a shear modulus about $2\times
10^{29}$ dyn $\rm cm^{-2}$ \citep{Ruderman91} in most of the crust
region of interest. For the magnetization $\pm M_{\rm S}$ of
Eq.(22), the strain resulting from the domain formation is
\bqn
 \widetilde{\varepsilon}&=&\pm
2\times10^{-7}(1-\frac{1}{a})^{1/2}(\frac{\epsilon}{100})^{-2}\nb\\
&& \times \left(\frac{G}{2\times
10^{29}}\right)^{-1}\left(\frac{B_{0}}{B_{\rm Q}}\right)^{3}.
\eqn 
The result of Eq.(31) shows that the magnetostrictive stain is
proportional to the third power of the central magnetic field in a
cycle if $a$ can be approximately treated as a constant and the
crust can bear before yielding. In spite of unknowing quantitatively
the limiting strain, the order of magnitude estimate of
$\widetilde{\varepsilon}_{\rm m}$ in the deep crust
\citep{Smoluchowski1970} is, $\widetilde{\varepsilon}_{\rm m}\sim
10^{-5}-10^{-3}$. The strain of Eq.(31), for a normal neutron star
with a magnetic field $B_{*}=0.1$ and $\psi_{0}=25$ MeV in Table-4, is
only about $8.0\times10^{-10}$. The normal neutron star can not be
cracked by the magnetostriction. The least magnetic field required
for cracking the deep outer crust is about (2.3-10.8)$\times B_{\rm
Q}$.

The magnetar magnetic field of $B_{0}\sim 10^{14}-10^{15}$G
\citep{Woods2006} inferred from the observational spin down of
magnetic braking is enough to crack the crust according to the
magnetostriction model. Illustrated as by the same case of a normal
neutron star above but with the magnetic field of a magnetar,
$B_{*}=10$, the strain is about $7.6\times10^{-4}$ in the range of
the limiting strain of $10^{-5}-10^{-3}$. This rupture will suddenly
release both of the magnetic and crustal elastic energies in the
form of the Alfven waves. Duncan and Thompson
\citep{DT92,thompson1995} had supposed that neutron stars with
fields beyond the quantum magnetic field of $B_{\rm Q}$ are the
sources of SGRs and AXPs bursts. Their proposed trigger mechanism of
bursts is also the solid crust fracture, but the origin is the
diffusing crustal magnetic field when built up sufficiently the
Maxwell stress. In fact, the sharply magnetizing of the electron gas
is an important or dominant contribution to the crustal fracture or
plastic flow. If the bursts energy derives truly from this part of
the free magnetic energy, the available maximum energy can be
approximated to (detailed computation see \cite{Wang2013}), $E_{\rm
m}=\frac{8\pi^{2}}{3}M_{\rm S}^{2}R_{*}^{3}$, with the neutron star
radius $R_{*}$. The maximum energy is the order of magnitude
$10^{38}(\frac{B_{0}}{B_{\rm Q}})^{4}$ erg as scaling $R_{*}$ to 10
km, $\epsilon$ to 100 and $a$ to $\infty$. Comparing the typical
repeat burst energy $E_{burst}<10^{41}$ ergs with the estimated
energy, the result indicates that the sudden crustal fracture and
the displacement of the magnetic footpoints driven by the
magnetostriction release enough energy to power the SGRs and the
AXPs repeat events. If the inferred dipolar magnetic field
$7.5\times10^{12}$ G  of SGR 0418+5729 \citep{Rea10} is the lowest
magnetic field magnetar, the least energy available for observations
is only about $10^{34}$ ergs. Such a low magnetic field does not
crack the neutron star crust. We need new burst models to explain
the low magnetic field magnetar (like-magnetar) emission.

\subsection{Barkhausen effects}

In Eq.(24), except ignoring the elastic energy of the solid crust we do
not consider the crystal disorders and defects as well. However, once these
are taken into account, an additional term should appear in the right-hand side of the
equation which is called quenched noise. This general equation
describing the motion of a driven interface in disordered medium
belongs to the KPZ equation \citep{Kardar1986}. Arising from the
quenched randomness, a pinned phase of the driven interface can exist
under the absence of an external field or it is lower than a threshold
value (determined by the disordering). As increasing the
driving field, the interface motion must experience a phase
transition from the pinned phase to a slowly smooth motion phase
interspersed with jumps. This is an example of the so-called
depinning phase transition.

The intermittent motion of DW in ferromagnetic materials (i.e., the
BK effect) has been studied extendedly by both experimental and
theoretical methods. All the essential elements of the BK effect as
discovered by experiments are the statistical properties of the
magnetization noise: distributions of duration and sizes of the
avalanches, etc. The BK noise is self-similar, and shows scaling
invariance or the power law distribution of statistical measure. In
other words, it has the typical features of a self-organized
criticality \citep{Aschwanden2014}. The exponent of the avalanche size
scales as a function of the constant change rate $C$ of the external field,
and fluctuation $D$ of the effective pinning field with Brownian
correlations \citep{Zapperi1998}. The result of \cite{Zapperi1998}
is
\begin{equation}
\tau=3/2-C/2D \mbox{ .}
\end{equation}

Behaving like all the first-order transitions, we expect that the
motion of DWs in the diamagnetic phase is similar to the Barkhausen
effect. It is natural to expect all the phenomena respecting to such a
diamagnetic phase, e.g., nucleation, hysteresis and supercooling.
The first hysteresis in the dHvA effect during the Condon domain formation
has been discovered in beryllium \citep{Kramer2005-1}. Actually, the
hysteresis is very small because of a high quality single crystals
demand and (or) perhaps the large thickness of DW
\citep{Egorov2010}. Such a very small Dingle temperature and a
relatively large rate of  the field can prevent DW pinning at lattice
point defects during wall movement in this experiment. However, the
situations of a neutron star with a very lowly magnetic field
evolution and high crystal disorder are significantly different from
the hysteresis test in beryllium.

If treating the interface of the domains of a neutron star as an
infinite-range elastic membrane with a fixed surface energy, and
simplifying the disordering as pointlike defects with Brownian
correlations we can expect the intermittent motion of DW with a
power-law avalanche size distribution. Supposing that the bursts of
SGR 1806-20 originate indeed from the depinning transition, the
observable power law index $\tau\approx 1.6$ \citep{cheng1996}
indicates that the  constant $C/D$ of Eq.(32) is about
-0.2. The minus sign means the decay of  the magnetic field. The non-small amplitude
implies that the long-range correlation of the effective pinning
field is not due to the internal correlation of the impurities.
Actually, these impurities are either uncorrelated or  only
short-range correlated. The real disorders determine the Ohmic
decay with a timescale larger than the Hall timescale at late times
of a neutron star. In this regime, the Hall effect dominates the
evolution of the crustal currents with a timescale depending strongly on
the internal field. For a typical crust of $Y_{\rm e}=0.25$,
$g_{14}=1$, the Hall timescale \citep{Cumming2004} has an
approximate value, $\tau_{\rm H}\sim 5.7\times10^{4}
\rho_{12}^{5/3}/B_{12}$ yrs ($B_{12}=B_{0}/10^{12}$). According to
the timescale, $\tau_{\rm H}$, and the fundamental frequency, $b$,
appearing in Eq.(13), a cycle of the dHvA magnetic oscillation actually
corresponds to a temporal period, $T_{\rm osci}\sim 0.25 \rho_{12}$
yrs. During this period, only a piece of the time associates with the
formation of the diamagnetic phase. In the outer crust, we can
expect that the jerky motion occurs frequently because of the short
cycle period and the many layers of the dHvA oscillations. The
outbursting emission observed in transient magnetars may be ascribed
to the continually occurring of many depinning transitions. In the
deep crust, however, not only the burst active seldom is  stronger than in that
 the outer crust, but also the BK
noise. If we take
$\rho_{12}=6$, the electron gas dominates the pressure, an expected
burst active period of 1.5 yrs between successive actives is roughly
in agreement with the observation of 2.4 yrs in SGR 1806-20
\citep{Laros1987}.

However, as in the usual first-order phase transition, if there is not
sufficient nucleation there may be a delay analogous to
supercooling. Deep in the Fermi liquid of the nucleus of a neutron star,
the domain structure does not appear because of the absence of the
disorders. Instead of the formation of the Condon domain, there will be a
metastable state corresponding to a homogeneous magnetization. As
the magnetized system evolves into the spinodals of both
homogeneous and  coexisting phases or undergoes a large
perturbation of the crust quake or BK noise, the loss balance of the
metastable state would accompany sharply variations of the internal
field which perhaps associates with the mechanism of giant flares
(energies in range $10^{44}-10^{46}$ ergs)
\citep{Mazets79,Hurley1999,Palmer05}.

\section{Conclusions}

We have applied the Lifshitz-Kosevich-Shoenberg formula to dense
matter with a relativistic electron gas. For a large Fermi energy
(having a large Landau quantum number), the oscillatory properties of
both no-relativistic and relativistic gases have the same form after the simple replacement
of the cyclotron mass by the relativistic mass. Moreover,
comparing with the no-relativistic gas, the magnetic interaction of
electrons of a neutron star is  strong enough to cause the diamagnetic
phase transition. At the low magnetic field end of the phase transition
curve, the phase-transition and the Dingle temperatures are
determined simply by the existence  of the
 dHvA oscillation, i.e., $\hbar\omega_{\rm c}\geq kT$ and
$\hbar\omega_{\rm c}\geq kT_{\rm D}$. Except the distinction of
interaction mechanisms, the diamagnetic phase transition is similar
to the ferromagnetic phase transition. By the same considerations as
in ferromagnetic materials, the special Condon domain configuration,
its size, and the surface tension of DW in a neutron star have been
calculated approximately.

In the observation of a neutron star, we suggested the
connection between the magnetar emission and the magnetostriction or the
Barkhausen effect. Due to the additional anisotropic stresses
associated with the diamagnetic phase, the solid crust of a magnetar
can be fractured which can trigger the observable X- or $\gamma-$
ray bursts. This is consistent with the radiation mechanism of
the Duncan and Thompson model. However, another self-organized
criticality of the BK effect can also explain the magnetar bursts. Here,
the driver of the self-organized system is the slowly and continuously
diffusive internal magnetic field due to the Hall effect and the ohmic
decay. The instability threshold is the pinning threshold which does
not reflect peculiar long-range correlations of the disorders, but
indicates an effective description of the collective motion of the
interface \citep{Cizeau1997}. Emergence of the magnetic flux by
magnetizing builds up the non-potential free magnetic energy that can be
released in the subsequent avalanches. This process of energy releasing
may be very efficient because of the damping of the long-range
correlatively effective pinning field acting on the Hall drift. The
mechanism makes the Hall effect itself become dissipative but be
distinct from the Hall cascade. With the new energy releasing
mechanism, we can explain the low-field magnetar bursts, the heating
mechanism of transient magnears, and the trigger mechanism of giant
flares, etc. However, more observational evidences and theoretical
comparisons  are needed to test the new mechanism.

\section*{ACKNOWLEDGMENTS}
We thank the anonymous referee for his/ her comments which helped to
improve the paper. ZJ thanks Prof. Anzhong Wang for correcting the
English language of the manuscript.This work was supported by
XinJiang Science Fund for Distinguished Young Scholars under Nos.
2014721015 and 2013721014, the National Science Foundation of China
(Grants Nos. 11503008, 11473024, 11363005, 11163005 and 11264037),
and the Doctor Foundation of Xinjiang University (Grant
No.BS110108).

\bibliographystyle{apj}
\bibliography{axp}

\end{document}